# К вопросу о классическом аналоге задачи Фано

М.В. Лебедев, О.В. Мисочко

Институт физики твёрдого тела РАН, 142432 Черноголовка, Московская область

Обсуждается система взаимодействующих классических осцилляторов, аналогичная, в некотором смысле, квантово-механической системе дискретного энергетического уровня, взаимодействующего с энергетическим квазиконтинуумом состояний, рассмотренной Фано. Анализируется предельный переход к случаю непрерывного спектра и возможная связь исследуемой задачи с генерацией когерентных фононов.

**Содержание**

1. **Введение**

Настоящая работа посвящена обсуждению попыток найти классическую механическую колебательную систему, которая при воздействии внешних периодических сил обладала бы спектральной зависимостью квадрата амплитуды колебаний совпадающей по виду с резонансным контуром Фано и состояла из классических элементов, каждый из которых являлся бы классическим аналогом соответствующего квантово-механического элемента модели Фано. Поясним сначала кратко, почему попытки такого рода представляют интерес с различных точек зрения.

**1.1 История вопроса**

Задача Фано возникла из проблемы теоретического описания характерных асимметричных линий поглощения благородных газов в ультрафиолетовой области спектра, возникающих на фоне широкого нерезонансного континуума поглощения [1]. Подход работы [1] был в дальнейшем успешно развит для описания рассеяния пучка электронов на атомах гелия [2]. Оказалось, что минимум в зависимости сечения рассеяния от энергии налетающих электронов, то есть подавление рассеяния, можно объяснить квантово-механическим эффектом деструктивной интерференции амплитуд электронной волны не поглощённой атомом и волны поглощённой с образованием метастабильного возбуждённого состояния атома, которое с течением времени распадается снова в непрерывный спектр состояний. Другими словами, состояния дискретного и непрерывного спектров могут влиять друг на друга и задачу можно сформулировать так: как изменятся состояния дискретного и непрерывного спектров при включении некоторого малого взаимодействия, переводящего частицу из состояния дискретного в состояние непрерывного спектра. Такая постановка вопроса оказалась настолько общей, что вскоре обнаружилось, что задача Фано возникает в самых разных областях физики. Было установлено, что появление асимметричных линий Фано это общее явление в таких различных областях исследований, как ядерная [3], атомная [4–6] и твердотельная физика [7–10], а также химия [11]. Резонансы Фано интересны своей общностью и до сих пор вызывают оживлённые дискуссии [12-14], а оригинальная работа [2] является сейчас одной из самых часто цитируемых в истории физики. Долгое время считалось, что интерференция Фано это чисто квантово-механическое явление, не имеющее классического аналога, поскольку при рассеянии классических частиц интерференционных явлений наблюдаться не может, однако спектральные контуры, напоминающие по виду контур Фано, возникают и при рассмотрении классических систем, представляемых механическими [15-17] или эквивалентными электрическими осцилляторами [18,19]. Именно внешнее сходство спектров классических и квантово-механических систем стимулировало поиски классической задачи аналогичной задаче Фано. Оказалось, что характерный асимметричный резонансный контур можно получить уже в системе всего лишь двух связанных между собой осцилляторов [15]. Принципиально важно при этом, чтобы один из осцилляторов обладал отличным от нуля затуханием. Несмотря на то, что рассмотрение [15] весьма поучительно, трудно признать систему из двух связанных осцилляторов аналогичной системе, состоящей из дискретного состояния, взаимодействующего с континуумом других состояний. Гораздо ближе к задаче Фано выглядит задача о классическом осцилляторе, связанном с квазиконтинуумом других классических осцилляторов, рассмотренная в работе [16]. В этой работе удалось получить контур Фано для классической системы очень похожей на исходную квантово-механическую, однако отметим, что задача Фано сформулирована в квантовой механике для консервативной системы, тогда как в классической аналогии [16] принципиальную роль играет затухание осцилляторов, которое нарушает симметрию по обращению времени и делает уравнения движения неконсервативными [20].

## 1.2 Физический смысл

Интерес к поиску связей между классическим и квантово-механическим описанием мира восходит к принципу соответствия Бора и задача Фано выглядит в этом отношении очень привлекательно. Рассмотрение классических аналогий квантово-механических явлений методически полезно и часто становилось предметом заметок УФН [21-23]. Ситуация, когда один осциллятор взаимодействует с ансамблем осцилляторов другой природы, очень часто встречается в физике твёрдого тела и при рассмотрении взаимодействия излучения с веществом. При этом многие коллективные возбуждения в твёрдых телах формируются в результате согласованного движения большого числа носителей заряда и, по сути, представляют собой квазиклассические состояния с достаточно хорошо определённой фазой колебаний. Интерес к поиску аналогий между задачей Фано и классическими колебательными системами был отчасти стимулирован последними достижениями в физике когерентных фононов – синхронного колебательного возбуждения кристаллической решётки в результате воздействия короткого лазерного импульса, длительность которого меньше периода колебаний атомов [24-26]. Спектры этих колебаний демонстрируют в определённых условиях характерный асимметричный контур, весьма напоминающий контур Фано [23]. Асимметричные профили линий спонтанного рамановского рассеяния в кристаллах наблюдались и ранее в условиях непрерывного лазерного возбуждения и трактовались как результат взаимодействия фонона с электронным континуумом [27]. Однако, эксперименты с временным разрешением позволили непосредственно измерять фазу колебаний атомов и естественным образом возник вопрос о связи этой фазы с механизмом возбуждения когерентных колебаний [16,17,19]. В работе [16] была предпринята попытка провести последовательное классическое рассмотрение задачи Фано, в которой модель Фано используется для объяснения зависимости начальной фазы когерентных фононов, возбуждаемых сверхкоротким световым импульсом в кристалле кремния, от уровня легирования этого кристалла донорными примесями [28]. Классические уравнения движения получаются при этом из квантового гамильтониана системы в результате предположения о том, что возбуждается когерентное состояние колебательной степени свободы кристалла, с которой слабо связаны электронные степени свободы, также находящиеся в когерентных состояниях. В результате когерентный фонон заменяется классическим гармоническим осциллятором, который предполагается слабо связанным с квазиконтинуумом классических гармонических осцилляторов, представляющих электронные степени свободы кристалла. Автору [16] удалось найти связь между начальной фазой когерентных колебаний решётки кристалла после возбуждения его сверхкоротким лазерным импульсом и параметром Фано спектральной линии данного колебания в спектре спонтанного комбинационного рассеяния света.

Применяя подход, развитый Фано, можно описать дисперсию квазичастиц, на что обратили внимание авторы работы [29] при рассмотрении дисперсии экситонных поляритонов в микрорезонаторе. При этом получается тот же результат, что и при использовании стандартного аппарата функций Грина. Оказалось, что анализ резонансного взаимодействия электромагнитной волны в таком микрорезонаторе с ансамблем двухуровневых систем, находящихся в случайном потенциале [30] приводит к уравнению на собственные значения энергии совпадающему с вековым уравнением для взаимодействующих классических осцилляторов, что прямо указывает на оправданность квазиклассического рассмотрения. Похожая ситуация возникает и при рассмотрении взаимодействия света с ансамблем атомов в задачах светового эха, когда энергия световой волны передаётся ансамблю локальных осцилляторов.

**1.3 Математические аспекты**

С математической точки зрения задача Фано сводится к отысканию собственных состояний квантово-механического гамильтониана, обладающего как дискретным, так и непрерывным спектром, при включении слабого резонансного взаимодействия между состояниями дискретного и непрерывного спектров [2]. Решением задачи являются суперпозиции исходных состояний дискретного и непрерывного спектров. При возбуждении таких состояний матричные элементы возбуждения дискретного уровня и континуума складываются с различными фазами, что приводит к возникновению характерной частотной зависимости возбуждения системы, называемой сейчас контуром Фано или интерференцией Фано, которая существенно отличается от Брейт-Вигнеровской зависимости (лоренцовская форма спектральной линии). Задача сводится, таким образом, к исследованию системы уравнений в частных производных. Что же касается упомянутых выше классических аналогий, то все они исследуют системы обыкновенных дифференциальных уравнений. Это обстоятельство будет обсуждаться ниже.

**1.4 Остающиеся проблемы**

Несмотря на внешнее сходство спектрального профиля классического и квантово-механического резонансов строгой аналогии между этими случаями до сих пор не установлено. Известные к настоящему моменту классические системы, демонстрирующие спектральную зависимость резонанса, напоминающую резонанс Фано, описываются гамильтонианами, не являющимися классическими аналогами квантово-механического гамильтониана задачи Фано. Внешнее сходство спектрального профиля резонанса, очевидно, не является достаточным основанием для проведения аналогии между квантовой и классической задачами. Попытки моделировать задачу Фано классическими осцилляторами, совершающими финитное движение, сталкиваются с проблемами, вытекающими из консервативности системы: воздействие гармонических сил на резонансной частоте такой системы приводит к неограниченному возрастанию амплитуды колебаний. Фаза колебаний осциллятора, возбуждаемого внешней гармонической силой, изменяется, как известно, скачком при переходе частоты возбуждающей силы через резонансную частоту. Получить плавное изменение фазы, как это имеет место в задаче Фано, можно только введя отличное от нуля затухание. В результате, в классических расчётах приходится рассматривать диссипативные системы, обладающие затуханием, тогда как квантово-механическая задача Фано затухания не содержит. Использование затухания оказывается принципиальным для классических расчётов, поскольку оно входит в выражения для параметра Фано. Переход к пределу нулевого затухания даёт симметричный лоренцевский спектральный профиль резонанса. Хотя строгой аналогии между задачей Фано и системой взаимодействующих классических осцилляторов, в силу названных причин, быть и не может, выяснение условий, при которых такая система демонстрирует спектральный профиль резонанса, напоминающий профиль Фано, представляет несомненный интерес.

В данной работе изложены результаты анализа движения системы связанных классических механических осцилляторов во многом эквивалентной рассмотренной в работе [16]. Новым, по сравнению с [16], является обсуждение свойств нормальных колебаний такой системы. Аналогом диагонализации квантово-механического гамильтониана для классической колебательной системы является нахождение её нормальных колебаний. Мы попытались проследить за нормальными колебаниями системы при неограниченном увеличении числа входящих в неё осцилляторов и получить предельное выражение для нормальных колебаний. Здесь возникает вопрос, связанный с неограниченным возрастанием амплитуды колебаний классической консервативной системы при её возбуждении на частотах нормальных колебаний. В выражениях, аналогичных по смыслу квантово-механическим, возникают в этом случае неустранимые расходимости и определить параметр Фано не удаётся. В работе [16] эта проблема была обойдена путём введения феноменологического затухания для осцилляторов, аналога которого не содержится в исходной задаче Фано. Отметим, что в работе Фано [1] был проведён расчёт для ограниченной квантовой системы, которая при возрастании её размера переходит в систему с дискретным уровнем и непрерывным спектром. Там аналогичные расходимости устраняются за счёт нормировки волновых функций, которая изменяется с ростом размеров системы. Физически понятно, что причиной расходимости в ограниченной консервативной системе при её резонансном возбуждении является сохранение энергии в такой системе, которая непрерывно возрастает в результате действия резонансной возбуждающей силы при отсутствии диссипации. Кажется разумным предположить, что при переходе к неограниченной системе, в которой энергия может уходить на бесконечность, должен появиться вариант решения проблемы упомянутых расходимостей без введения дополнительного затухания. Мы обсуждаем эту возможность ниже, после подробного анализа классической модели взаимодействующих локальных осцилляторов и показываем, что последняя не может являться аналогом задачи Фано при сколь угодно высокой спектральной плотности осцилляторов, образующих квазиконтинуум.

## 2. Модель связанных механических осцилляторов

Мы начинаем с описания модели связанных классических гармонических осцилляторов, которая аналогична модели рассмотренной в работе [16], но не содержит затухания. Модель в работе [16] используется для объяснения зависимости начальной фазы когерентных фононов от механизма их возбуждения. Мы также будем обсуждать применимость нашей модели для описания свойств когерентных фононов по мере выяснения особенностей её поведения под действием внешних сил и сравнивать наши результаты с результатами работы [16].

Рассмотрим механическую систему, изображённую на Рис.1. Тележка массы $M$ связана пружиной жёсткости $K$ с неподвижной стенкой и может совершать одномерное движение вдоль направляющих. Посредством слабых пружин с жесткостью $\mu_i$ эта тележка связана с набором $2N+1$ пружинных маятников, закреплённых на противоположной неподвижной стенке (массы шариков $m_i$, жёсткость пружин $k_i$).

Выражения для кинетической и потенциальной энергии такой системы имеют вид:

$$T = \frac{1}{2}M\dot{X}^2 + \frac{1}{2}\sum_{i=-N}^{N} m_i \dot{x}_i^2$$

$$\Pi = \frac{1}{2}KX^2 + \frac{1}{2}\sum_{i=-N}^{N} k_i x_i^2 + \frac{1}{2}\sum_{i=-N}^{N} \mu_i(x_i - X)^2 = \frac{1}{2}(K + \sum_{i=-N}^{N}\mu_i)X^2 + \frac{1}{2}\sum_{i=-N}^{N}(k_i + \mu_i)x_i^2 - \sum_{i=-N}^{N}\mu_i x_i X \qquad (1)$$

Отсюда следуют уравнения Лагранжа для движения системы:

$$m_N \ddot{x}_N + (k_N + \mu_N)x_N - \mu_N X = 0$$

$$m_{N-1}\ddot{x}_{N-1} + (k_{N-1} + \mu_{N-1})x_{N-1} - \mu_{N-1}X = 0$$

$$........$$

$$m_0\ddot{x}_0 + (k_0 + \mu_0)x_0 - \mu_0 X = 0 \qquad (2)$$

$$M\ddot{X} + (K + \sum_{i=-N}^{N}\mu_i)X - \sum_{i=-N}^{N}\mu_i x_i = 0$$

$$......$$

$$m_{-N}\ddot{x}_{-N} + (k_{-N} + \mu_{-N})x_{-N} - \mu_{-N}X = 0$$

Для нахождения нормальных колебаний, запишем координаты пружинных маятников и тележки в виде:

$$x_i = u_i \sin(\omega t + \alpha)$$

$$X = U \sin(\omega t + \alpha) \qquad (3)$$

Величины $u_i$ и $U$ представляют собой амплитуды смещений отдельных пружинных маятников и тележки, соответственно. Считается, что амплитуда может быть как положительной, так и отрицательной, поэтому фаза нормального колебания $\alpha$ определена на интервале $(0, \pi)$. Подставляя выражения (3) в уравнения движения (2), получаем систему алгебраических уравнений для нахождения амплитуд нормальных колебаний:

$$(k_N + \mu_N - m_N \lambda)u_N - \mu_N U = 0$$

……

$$(K + \sum_{i=-N}^{N} \mu_i - \lambda M)U - \sum_{i=-N}^{N} \mu_i u_i = 0 \qquad (4)$$

……..

$$(k_{-N} + \mu_{-N} - \lambda m_{-N})u_{-N} - \mu_{-N} U = 0$$

Введём обозначения:

$$\frac{k_i + \mu_i}{m_i} = \omega_i^2 \qquad \frac{\mu_i}{m_i} = c_i \qquad \frac{K + \sum_{i=-N}^{N} \mu_i}{M} = \Omega^2$$

Тогда система уравнений (4) примет вид:

$$(\omega_N^2 - \lambda)u_N - c_N U = 0$$

……….

$$(\Omega^2 - \lambda)U - \sum_{i=-N}^{N} \frac{m_i}{M} c_i u_i = 0 \qquad (5)$$

……….

$$(\omega_{-N}^2 - \lambda)u_{-N} - c_{-N} U = 0$$

Ненулевые решения этой системы уравнений существуют только при условии равенства нулю её определителя, которое даёт вековое уравнение для нашей системы осцилляторов:

$$\Omega^2 - \lambda = \sum_{i=-N}^{N} \frac{m_i}{M} \frac{c_i^2}{\omega_i^2 - \lambda} \qquad (6)$$

Это алгебраическое уравнение относительно $\lambda$ порядка $2N + 2$, равного числу степеней свободы, корни которого определяют нормальные колебания нашей системы.

Перейдём к исследованию векового уравнения (6). До сих пор мы не делали никаких предположений о собственных частотах и связях между осцилляторами. Поскольку наше рассмотрение имеет целью дать классический аналог взаимодействий элементарных возбуждений в твёрдых телах, то чтобы такая аналогия имела смысл, необходимо предположить, что жесткости пружин, связывающих пружинные маятники с тележкой малы в сравнении с пружинами осцилляторов $\mu_i \ll k_i, K$. «Перенормировка» исходных собственных частот осцилляторов, вызванная взаимодействием, будет в этом случае невелика:

$$\frac{k_i+\mu_i}{m_i} = \omega_i^2 \approx \omega_{i0}^2 = \frac{k_i}{m_i} \qquad (7)$$

Для тележки это условие приводит к более сильному ограничению, поскольку требуется, чтобы суммарная жесткость всех связей была мала по сравнению с жесткостью пружины тележки:

$$\sum_{i=-N}^{N} \mu_i \ll K \qquad (8)$$

Заметим, что частоты, фигурирующие в вековом уравнении (6), представляют собой «перенормированные» частоты исходных осцилляторов, то есть изменённые за счёт дополнительных упругих связей между тележкой и осцилляторами. Говоря в дальнейшем о частотах осцилляторов, мы всегда будем иметь в виду именно эти «перенормированные» частоты.

3. **Исследование векового уравнения для двух связанных механических осцилляторов**

Рассмотрим графическое решение векового уравнения, которое даёт наглядный способ качественного анализа свойств рассматриваемой системы связанных осцилляторов. Пусть мы имеем сначала всего один осциллятор, связанный с тележкой ($N = 0, i = 0$). Тогда левая часть векового уравнения представляет собой линейную, а правая часть – гиперболическую функцию переменной $\lambda$.

$$\Omega^2 - \lambda = \frac{m_0}{M} \frac{c_0^2}{\omega_0^2 - \lambda} \qquad (9)$$

Случай, когда частоты совпадают, показан на Рис.2а. Точки пересечения графиков левой и правой частей векового уравнения определяют частоты нормальных колебаний системы. Уравнения (5) выглядят при этом следующим образом:

$$(\omega_0^2 - \lambda)u_0 - c_0 U = 0$$
$$(\Omega^2 - \lambda)U - \frac{m_0}{M} c_0 u_0 = 0 \qquad (10)$$

Из них непосредственно видно, что при $\lambda < \Omega^2$ знаки амплитуд $U$ и $u_0$ совпадают, то есть тележка и пружинный маятник движутся синфазно. В случае же $\lambda > \Omega^2$ их движения противофазны. Два решения векового уравнения дают полный набор нормальных колебаний, так что любое движение может быть представлено как их суперпозиция. Уже эта простейшая модель связанных осцилляторов часто используется для описания возникновения поляритонов – элементарных возбуждений в задаче о поляризационных колебаниях среды, сильно взаимодействующих с электромагнитным полем [12]. Интересно, что наша механическая система позволяет при этом дать наглядное представление о частоте Раби. В самом деле, пусть в начальный момент мы отклоняем только один из осцилляторов, например, пружинный маятник, в то время как второй (тележка) покоится. Это начальное состояние является, очевидно, суперпозицией синфазного и противофазного колебаний системы, входящих с одинаковыми по модулю коэффициентами. Дальнейшее движение системы будет представлять собой биения с частотой, равной разности частот нормальных колебаний, в процессе которых вся энергия периодически будет переходить от пружинного маятника к тележке и обратно. Частота обмена энергией между осцилляторами равная разности частот нормальных колебаний и будет механическим аналогом частоты Раби [13]. Непосредственно из уравнений (10) при этом следует:

$$(\Omega^2 - \lambda) = \pm c_0 \sqrt{\frac{m_0}{M}} \qquad (11)$$

Откуда для частот нормальных колебаний $\omega_+$ и $\omega_-$ и частоты Раби $\Omega_R$ получаем:

$$\omega_+ = \sqrt{\Omega^2 + c_0\sqrt{\frac{m_0}{M}}} \approx \Omega + \frac{1}{2}\sqrt{\frac{m_0}{M}}\frac{c_0}{\Omega}$$

$$\omega_- = \sqrt{\Omega^2 - c_0\sqrt{\frac{m_0}{M}}} \approx \Omega - \frac{1}{2}\sqrt{\frac{m_0}{M}}\frac{c_0}{\Omega}$$

$$\Omega_R = \omega_+ - \omega_- = \sqrt{\frac{m_0}{M}}\frac{c_0}{\Omega} = \frac{1}{\sqrt{m_0 M}}\frac{\mu_0}{\Omega} \qquad (12)$$

Видно, что частота Раби прямо пропорциональна упругости пружины, связывающей тележку и пружинный маятник.

Случай, когда частоты колебаний тележки и пружинного маятника различны, показан на Рис.2b. Новым, по сравнению с предыдущим случаем, является преобладание свойств одного из маятников в конкретном нормальном колебании системы. В самом деле, частоты нормальных колебаний, как видно из Рис.2b, приближаются, по мере увеличения разности частот осцилляторов, к частотам соответствующих осцилляторов, при этом амплитуда колебаний того из осцилляторов, к частоте которого приближается частота нормального колебания доминирует, как это видно из уравнений (10). Тем не менее, осцилляторы, как и раньше, движутся в нормальном колебании с меньшей частотой синфазно, а в колебании с большей частотой противофазно, так что, подобрав соответствующую суперпозицию, можно добиться того, чтобы один из маятников периодически полностью передавал свою энергию второму (то есть прекращал движение), но второй осциллятор при этом будет обмениваться только частью запасённой в нём энергии.

**4. Произвольное количество осцилляторов, вырожденный случай.**

Обратимся теперь к случаю многих взаимодействующих с тележкой осцилляторов. Начнём с вырожденной ситуации, когда частоты всех осцилляторов совпадают между собой, но не обязательно совпадают с частотой тележки. Примем, для простоты, что все осцилляторы и их связи с тележкой идентичны. Вековое уравнение принимает в этом случае вид:

$$\Omega^2 - \lambda = \frac{1}{\omega_0^2 - \lambda} \sum_{i=-N}^{N} \frac{m_i}{M} c_i^2 = (2N+1)\frac{1}{\omega_0^2 - \lambda}\frac{m_0}{M} c_0^2 \qquad (13)$$

Это уравнение с точностью до обозначений совпадает с уравнением (9), поэтому все выводы, сделанные для двух осцилляторов, остаются в силе и для вырожденного случая. Физически также понятно, что большое количество пружинных осцилляторов, частоты которых совпадают, можно заставить колебаться синхронно между собой. При этом, с точки зрения взаимодействия с тележкой совокупность этих маятников ведёт себя как один маятник, масса которого равна сумме масс отдельных осцилляторов, а жесткость – сумме жесткостей пружин этих осцилляторов. В случае совпадения частот осцилляторов с частотой собственных колебаний тележки можно говорить о коллективной частоте Раби [31]. Будем называть эти два нормальных колебания рассматриваемой системы, аналогичные нормальным колебаниям в системе двух осцилляторов, граничными нормальными колебаниями, в том смысле, что они определяют, как это будет ясно из дальнейшего, верхнюю и нижнюю границы спектра нормальных колебаний. Весь коллектив пружинных осцилляторов, связанных с тележкой, выступает при этом как один эффективный осциллятор. Наша механическая аналогия хорошо описывает случай многоатомных Раби осцилляций в системе одинаковых двухуровневых атомов, взаимодействующих с резонансным электромагнитным полем [30]. Коллективная частота Раби пропорциональна при этом корню из общего числа осцилляторов. Работа [30] интересна тем, что в ней рассматривается общий случай резонансного взаимодействия электромагнитного поля с невырожденным ансамблем двухуровневых систем, частоты которых расположены в некоторой окрестности частоты электромагнитной волны. Уравнение на собственные значения энергий, полученное в [30], с точностью до обозначений, совпадает с вековым уравнением (6) и поведение нашей механической системы, как будет ясно из дальнейшего рассмотрения, вполне аналогично результатам для системы квантово-механических осцилляторов.

Вернёмся к рассмотрению вырожденной системы осцилляторов. Что же представляют собой оставшиеся $2N$ нормальных колебаний системы? Нетрудно увидеть, что остальные нетривиальные решения системы уравнений (5) отвечают условию:

$$U = 0; \quad \sum_{i=-N}^{N} \frac{m_i}{M} c_i\, u_i = 0 \tag{14}$$

Последнее условие есть не что иное, как условие ортогональности векторов в пространстве $2N + 1$ измерений, которое определяет $2N$-мерную плоскость в этом пространстве. Число независимых взаимно ортогональных векторов, задающих эту плоскость равно как раз $2N$, что совпадает с числом недостающих нормальных колебаний системы. Физически эти колебания происходят таким образом, чтобы результирующая сила, действующая на тележку, всегда была строго равна нулю. В результате вся колебательная энергия, содержащаяся в системе, сосредоточена в ансамбле пружинных осцилляторов и обмена ею с тележкой не происходит. Частота колебаний при этом вырождена и совпадает с общей частотой всех пружинных осцилляторов (на графике этим решениям отвечает точка $\lambda = \omega_0^2$). Это можно увидеть, рассматривая определитель системы уравнений (5). При одинаковых осцилляторах все диагональные элементы этого определителя за исключением одного (соответствующего амплитуде колебаний тележки) одновременно обращаются в нуль. Раскладывая определитель по строкам можно убедиться, что он равен нулю, если хотя бы два его диагональных элемента одновременно равны нулю. Это означает, что если имеется хотя бы два пружинных осциллятора с совпадающими собственными частотами, то их общая частота становится частотой нормального колебания системы. Действительно, эти два осциллятора могут, очевидно, колебаться в противофазе друг с другом так, чтобы суммарная сила, действующая на тележку, всегда была равна нулю. Амплитуды колебаний всех остальных осцилляторов и тележки можно положить равными нулю, при этом система уравнений (5) будет удовлетворяться.

### 5. Произвольное количество осцилляторов, снятие вырождения.

Проследим теперь за изменениями нормальных колебаний системы осцилляторов при снятии вырождения. Наглядный качественный анализ происходящего можно провести, рассматривая графическое решение векового уравнения. Пусть собственная частота $\omega_0$ одного из осцилляторов отличается от общей собственной частоты остальных $2N$ осцилляторов $\omega_N$ и собственной частоты колебаний тележки $\Omega$, тогда вековое уравнение примет вид:

$$\Omega^2 - \lambda = \frac{1}{\omega_0^2 - \lambda} \sum_{i=-N}^{N-1} \frac{m_i}{M} c_i^2 + \frac{m_N}{M} c_N^2 \frac{1}{\omega_N^2 - \lambda} \qquad (15)$$

Появление нового полюса в правой части приводит к дополнительному разрыву в её графике и к дополнительной точке пересечения с линейной функцией, представляющей левую часть (см. Рис.3a). При этом число вырожденных решений уменьшается на единицу. Пусть теперь все частоты пружинных осцилляторов различны и сосредоточены в некотором интервале вокруг частоты тележки. Ни одна из собственных частот осцилляторов теперь уже не будет совпадать с частотой какого-либо нормального колебания системы, а в графике правой части векового уравнения будет существовать $2N + 1$ разрывов и будет иметься $2N+2$ точки пересечения этого графика с линейной функцией левой части (см. Рис.3b). Знак и величина амплитуды $i$-го осциллятора в $j$- ом нормальном колебании следует из соотношения (5):

$$(\omega_i^2 - \lambda_j)u_i = c_i U \qquad (16)$$

Во всех нормальных колебаниях с $\lambda_j < \omega_i^2$ он колеблется синфазно с тележкой, а в нормальных колебаниях с $\lambda_j > \omega_i^2$ в противофазе с ней (см. Рис.4). Нормальное колебание с наименьшей частотой будет являться синфазным граничным колебанием, а с наибольшей частотой – противофазным граничным колебанием ансамбля осцилляторов и тележки, что схематически иллюстрируется Рис. 5.

Мы видим, что, хотя принципиальных отличий между граничными колебаниями в вырожденной и невырожденной системе нет, важная особенность вырожденной (или близкой к вырождению) системы заключается в том, что амплитуды отдельных осцилляторов в граничных нормальных колебаниях в такой системе могут быть близки, а это означает возможность почти полного обмена энергией между тележкой и пружинными маятниками в ходе коллективных осцилляций Раби. Принципиальным отличием вырожденной системы является также существование нормальных колебаний, в которых суммарное взаимодействие осцилляторов с тележкой равно нулю.

## 6. Поведение системы под действием внешних сил.

Обратимся теперь к рассмотрению поведения нашей системы при её различном возбуждении. Можно ли моделировать таким образом механизмы возбуждения когерентных фононов? В физике когерентных фононов вследствие прямого измерения разрешённых во времени колебаний кристаллической решётки появляется новая экспериментально измеряемая величина – начальная фаза когерентных колебаний. Хорошо установлено, что начальная фаза различна для различных кристаллов и её принято связывать с механизмом возбуждения когерентных фононов. Эта начальная фаза зависит по существу от положения равновесия колеблющихся атомов. При возбуждении когерентных фононов происходит кратковременное воздействие на осцилляторы импульсной силы. Действие на систему, находящуюся в равновесии, короткой импульсной силы (продолжительность воздействия много меньше любого из периодов колебаний маятников) можно заменить введением начального условия с отличными от нуля начальными скоростями и рассматривать дальше свободную эволюцию такой системы.

Ясно, что после окончания лазерного импульса система начинает движение из положения равновесия и начальная фаза колебаний может равняться либо 0 либо π. Такая начальная фаза действительно наблюдается во многих прозрачных на частоте лазерного возбуждения кристаллах, в которых ширина энергетической щели превышает энергию возбуждающих фотонов [23,26]. Это возбуждение когерентных фононов принято называть динамическим и связывать с динамическим рамановским механизмом возбуждения. В непрозрачных кристаллах когерентные колебания, напротив, начинаются обычно из положения максимальной амплитуды, то есть с начальной фазой равной $\frac{\pi}{2}$ или $-\frac{\pi}{2}$. В этом случае, в результате действия лазерного импульса происходит мгновенное изменение положения равновесия атомов. Такое изменение положения равновесия можно описать параметрическим механизмом возбуждения, когда за счёт действия импульса мгновенно смещается на конечную величину стенка, за которую закреплены пружины осцилляторов или мгновенно изменяется жёсткость пружин. Это соответствует мгновенному изменению межатомного потенциала в кристалле, которое принято связывать с высокой плотностью фотовозбуждённых электронов и экранировкой межатомного взаимодействия. Атомы начинают в этом случае колебания с нулевой начальной скоростью и такое возбуждение принято называть кинематическим. Если сравнить спектры Фурье когерентных колебаний при динамическом и кинематическом возбуждении (кинематическое или динамическое по отношению к осциллятору с дискретным спектром), то спектр динамического возбуждения будет содержать только частоту возбуждаемого фонона, в то время как спектр кинематического возбуждения будет содержать ещё и континуум частот, вызванный скачкообразным появлением конечной амплитуды. Соотношение между дискретной частотой и континуумом в спектре Фурье когерентных колебаний должно зависеть, таким образом, от начальной фазы колебаний. Интересно, в этой связи, несколько иначе взглянуть на формулу Фано. Если спектр, отвечающий формуле Фано, является спектром некоторого когерентного колебания, то временной ход этого колебания можно определить, выполнив обратное преобразование Фурье контура Фано. Такое преобразование было выполнено нами в работе [19]. Результатом этого преобразования являются колебания затухающего гармонического осциллятора с собственной частотой $\Omega$ , начальная фаза колебаний которого действительно определяется параметром Фано.

Решив вековое уравнение, можно определить частоты всех нормальных колебаний связанных осцилляторов и разложить любое начальное состояние системы, возникшее в результате импульсного воздействия, по её нормальным колебаниям. Воздействие сверхкороткого импульса можно моделировать, как сказано выше, соответствующим начальным условием: действие импульсной силы на тележку заменить сообщением ей в нулевой момент времени некоторой начальной скорости, действие же импульса возбуждения на пружинные осцилляторы описать, задав в нулевой момент времени их начальное отклонение от положения равновесия. Дальнейшую эволюцию системы легко рассчитать на основании разложения по нормальным колебаниям.

Нами было выполнено численное моделирование нашей системы на компьютере с количеством пружинных маятников равным 200. После задания собственных частот и связей производилось численное решение векового уравнения и определялась эволюция системы с тем или иным начальным условием. Оказалось, что получить затухающие колебания тележки с различной начальной фазой, как можно было бы ожидать на основе модели возбуждения когерентных фононов, не удаётся. Если в начальный момент были возбуждены колебания тележки, то вначале их амплитуда действительно начинает уменьшаться из-за передачи энергии к пружинным осцилляторам и быстрой дефазировки их колебаний, как это обычно и предполагается в физической картине светового эха. В дальнейшем, однако, наблюдается и обратная передача энергии от пружинных осцилляторов к тележке, так что говорить об экспоненциальном затухании её колебаний со временем не приходится. Оставляя в стороне тонкие вопросы о точности измерения частоты, погрешностях задания начальных условий и связанной с этим общей предсказуемости поведения детерминированной системы, заметим лишь, что в общем случае обмен энергией между тележкой и осцилляторами может оказаться достаточно нетривиальным во времени. Количество осцилляторов в нашем численном моделировании, разумеется, очень мало по сравнению с физически интересными ситуациями, когда их количество может достигать величин порядка числа Авогадро, т.е. $10^{23}$. Можно, поэтому, надеяться, что при очень большом числе осцилляторов проявится экспоненциальное затухание колебаний в соответствии с Фурье-преобразованием формулы Фано. Другая возможность обеспечить экспоненциальное затухание амплитуды когерентных колебаний – введение затухания для пружинных осцилляторов.

В некоторых случаях детального знания возбуждённых нормальных колебаний не требуется. Очевидно, например, что при импульсном возбуждении с шириной спектра возбуждающей силы большем полной ширины спектра нормальных колебаний, будут возбуждены практически все нормальные колебания системы. Эволюция после такого импульсного воздействия будет существенно зависеть от соотношения частот нормальных колебаний. В самом деле, представим себе, что все частоты нормальных колебаний кратны некоторой основной частоте, как это бывает для колебаний в резонаторах. Тогда начальное состояние, возникшее сразу после окончания возбуждения, будет периодически повторяться. Если импульсное воздействие осуществлялось только на тележку, то энергия в процессе эволюции будет передаваться сначала пружинным осцилляторам, но через время, равное периоду колебания основной частоты, которой кратны все частоты нормальных колебаний, пружинные осцилляторы вновь окажутся неподвижными и вся энергия вновь сосредоточиться в колебании тележки. Можно показать, что такая основная частота всегда существует, если отношение частот нормальных колебаний выражается рациональными числами.

**6.1 Воздействие на систему гармонических сил**

Будем считать систему осцилляторов невырожденной и пусть на неё действуют гармонические вынуждающие силы:

$$f_i(t) = f_i(\omega) \sin(\omega t + \alpha_{f_i})$$

$$F(t) = F(\omega) \sin(\omega t + \alpha_F) \qquad (17)$$

Сосредоточимся только на вынужденном движении системы, и будем искать решения также в виде гармонических функций:

$$x_i = u_i(\omega) \sin(\omega t + \alpha_i)$$

$$X = U(\omega) \sin(\omega t + \alpha) \qquad (18)$$

Как и в случае поиска нормальных колебаний, нетрудно видеть, что при равенстве фаз действующих сил $\alpha_{f_i} = \alpha_F$ возможны только синфазные и противофазные движения осцилляторов, которые отражаются на знаках соответствующих амплитуд, то есть:

$$\alpha_i = \alpha = \alpha_{f_i} = \alpha_F$$

При этом мы получаем систему алгебраических уравнений:

$$(\omega_i^2 - \omega^2)u_i - c_i U = \frac{f_i}{m_i} \qquad i = -N, \ldots, N$$

$$(\Omega^2 - \omega^2)U - \sum_{i=-N}^{N} \frac{m_i}{M} c_i u_i = \frac{F}{M} \qquad (19)$$

## 6.2 Сила действует только на тележку

Рассмотрим сначала случай, когда все $f_i = 0$, то есть сила действует только на тележку. Находим:

$$u_i = \frac{c_i U}{\omega_i^2 - \omega^2}$$

$$U = \frac{F}{M} \frac{1}{(\Omega^2 - \omega^2) - \sum_{i=-N}^{N} \frac{m_i}{M} \frac{c_i^2}{\omega_i^2 - \omega^2}} \qquad (20)$$

Уравнение для нулей знаменателя выражения для амплитуды колебаний тележки совпадает с вековым уравнением. Эта амплитуда резонансно возрастает при приближении частоты возбуждающей силы к частотам нормальных колебаний системы. Интересно проследить за амплитудами тележки и $i$ −го осциллятора при изменении частоты вынуждающей силы в окрестности резонансной частоты осциллятора $\omega_i$ включающей частоты двух соседних нормальных колебаний $\omega_- = \omega_i - \delta_L$ и $\omega_+ = \omega_i + \delta_H$ (см. рис.6). Амплитуда колебаний осциллятора согласно уравнению (20) равна:

$$u_i = \frac{F}{M} \frac{c_i}{(\omega_i^2 - \omega^2)\left[(\Omega^2 - \omega^2) - \sum_{i=-N}^{N} \frac{m_i}{M} \frac{c_i^2}{\omega_i^2 - \omega^2}\right]} \qquad (21)$$

Она, также как и амплитуда колебаний тележки, резонансно возрастает при приближении частоты вынуждающей силы к частотам нормальных колебаний, причём фаза движения осциллятора по отношению к вынуждающей силе меняет знак при переходе частоты возбуждения через частоту осциллятора. Амплитуда колебаний тележки на частоте $\omega_i$ обращается в ноль, тогда как амплитуда колебаний осциллятора остаётся конечной и равной $u_i = -\frac{F}{m_i c_i}$, что очевидно, если переписать формулу (21) в виде:

$$u_i = \frac{F}{M} \frac{c_i}{(\omega_i^2 - \omega^2)\left[(\Omega^2 - \omega^2) - \sum_{k \neq i} \frac{m_k}{M} \frac{c_k^2}{\omega_k^2 - \omega^2}\right] - \frac{m_i}{M} c_i^2} \qquad (22)$$

Этот эффект в системе двух осцилляторов обсуждался в работе [15] в связи с резонансом Фано. Обращение амплитуды колебаний первого из двух связанных между собой осцилляторов в нуль при определённой частоте возбуждающей его гармонической силы объяснялось противофазным воздействием на него второго осциллятора. Данный результат согласуется с картиной, представленной на рис.6, однако не имеет, по нашему мнению, прямого отношения к задаче Фано.

**6.3 Сила действует только на один из пружинных маятников**

Пусть теперь отлична от нуля только сила, действующая на один из пружинных маятников. Уравнения принимают в этом случае следующий вид:

$$(\omega_i^2 - \omega^2)u_i - c_i U = \frac{f_i}{m_i}$$

$$(\omega_j^2 - \omega^2)u_j - c_j U = 0 \qquad j \neq i$$

$$(\Omega^2 - \omega^2)U - \sum_{k=-N}^{N}\frac{m_k}{M}c_k u_k = 0 \qquad (23)$$

Умножив каждое из уравнений для осцилляторов на $\frac{m_k}{M}c_k\frac{1}{(\omega_k^2-\omega^2)}$ и сложив эти уравнения с уравнением для амплитуды колебаний тележки, получаем:

$$U\left[(\Omega^2-\omega^2)-\sum_{k=-N}^{N}\frac{m_k}{M}\frac{c_k^2}{\omega_k^2-\omega^2}\right] = \frac{f_i}{M}\frac{c_i}{\omega_i^2-\omega^2} \qquad (24)$$

Откуда следует, что амплитуда колебаний тележки:

$$U = \frac{f_i}{M}\frac{c_i}{(\omega_i^2-\omega^2)\left[(\Omega^2-\omega^2)-\sum_{k=-N}^{N}\frac{m_k}{M}\frac{c_k^2}{\omega_k^2-\omega^2}\right]} \qquad (25)$$

Для амплитуд колебаний осцилляторов имеем:

$$u_j = \frac{f_i}{M}\frac{c_j c_i}{(\omega_j^2-\omega^2)(\omega_i^2-\omega^2)\left[(\Omega^2-\omega^2)-\sum_{k=-N}^{N}\frac{m_k}{M}\frac{c_k^2}{\omega_k^2-\omega^2}\right]} \qquad j \neq i \qquad (26)$$

$$u_i = \frac{f_i}{M}\frac{c_i^2}{(\omega_i^2-\omega^2)^2\left[(\Omega^2-\omega^2)-\sum_{k=-N}^{N}\frac{m_k}{M}\frac{c_k^2}{\omega_k^2-\omega^2}\right]} + \frac{f_i}{m_i}\frac{1}{\omega_i^2-\omega^2} \qquad (27)$$

Все амплитуды имеют резонансы на частотах нормальных колебаний системы и формула (25) для амплитуды колебаний тележки совпадает с точностью до замены $F$ на $f_i$ с формулой (21) для амплитуды колебаний осцилляторов в предыдущем случае. Поведение всех амплитуд при изменении частоты возбуждения в окрестности частот осцилляторов $\omega_i$ и $\omega_j$  $j \neq i$ показано на рис.7, где $\omega_+$ и $\omega_-$ обозначают, как и раньше две частоты нормальных колебаний соседние с частотой возбуждаемого осциллятора.

Рассмотрим подробнее поведение осцилляторов и тележки. Амплитуда колебаний тележки имеет резонансы на частотах нормальных колебаний системы. Вблизи частот нормальных колебаний, частота которых меньших частоты $\omega_i$ осциллятора возбуждаемого внешней силой, тележка движется в фазе с возбуждающей силой с низкочастотной стороны от резонанса и в противофазе с ней с высокочастотной стороны. Для частот нормальных колебаний больших $\omega_i$ фазы движения тележки меняются на обратные. На частоте $\omega_i$ тележка движется в противофазе с возбуждающей силой, амплитуда её колебаний конечна и равна:

$$U(\omega_i) = -\frac{f_i}{m_i c_i} \tag{28}$$

Амплитуда колебаний тележки обращается в нуль на частотах остальных осцилляторов:

$$U(\omega_j) = 0 \qquad j \neq i \tag{29}$$

Возбуждаемый внешней силой осциллятор также имеет резонансы на всех частотах нормальных колебаний, однако фаза его движения вблизи этих резонансов всегда совпадает с фазой вынуждающей силы с низкочастотной стороны и противоположна ей с высокочастотной стороны от резонанса. Амплитуда этого осциллятора обращается в нуль на частотах, задаваемых уравнением, следующим непосредственно из выражения (27):

$$M(\omega_i^2 - \omega^2)^2 \left[(\Omega^2 - \omega^2) - \sum_{k=-N}^{N} \frac{m_k}{M} \frac{c_k^2}{\omega_k^2 - \omega^2}\right] = -m_i c_i^2 (\omega_i^2 - \omega^2) \tag{30}$$

Помимо особой точки $\omega = \omega_i$, имеются дополнительные решения, удовлетворяющие уравнению:

$$(\Omega^2 - \omega^2) - \sum_{\substack{k=-N \\ k \neq i}}^{N} \frac{m_k}{M} \frac{c_k^2}{\omega_k^2 - \omega^2} = 0 \tag{31}$$

Это уравнение совпадает с вековым уравнением для системы осцилляторов, полученной из исходной исключением из неё осциллятора с частотой $\omega_i$. Один из корней этого уравнения обозначен на рис.7 как $\omega_z$. Обращение $u_i$ в нуль на частотах нормальных колебаний «укороченной» системы осцилляторов согласуется с аналогичным результатом предыдущего случая, когда сила была приложена к тележке. «Укороченная» система выступает по отношению к возбуждаемому внешней силой осциллятору как отдельная колебательная система со своими резонансными частотами, с которой взаимодействует данный осциллятор. Обращение $u_i$ в нуль на частотах нормальных колебаний «укороченной» системы совершенно аналогично, поэтому обращению в предыдущем случае в нуль амплитуды тележки на частотах нормальных колебаний исходной системы. В особой точке $\omega_i$ амплитуда $u_i$ имеет конечное значение. Это проще всего увидеть непосредственно из исходных уравнений (23). Подставляя $\omega = \omega_i$ в первое из уравнений, находим $U = -\frac{f_i}{m_i c_i}$. Этим определяются амплитуды остальных осцилляторов: $u_j(\omega_i) = -\frac{f_i c_j}{m_i c_i}\frac{1}{\omega_j^2 - \omega_i^2}$ $j \neq i$ и из последнего уравнения получаем:

$$u_i(\omega_i) = -\frac{Mf_i}{(m_i c_i)^2}\left[(\Omega^2 - \omega_i^2) - \sum_{k \neq i}\frac{m_k}{M}\frac{c_k^2}{\omega_k^2 - \omega_i^2}\right] \qquad (32)$$

Это значение может быть как положительным, так и отрицательным, а также и нулём, если частота $\omega_i$ окажется корнем «укороченного» векового уравнения (31). То, что такое возможно, иллюстрируется рис.8.

Вопрос можно переформулировать следующим образом: может ли собственная частота осциллятора стать частотой нормальных колебаний после удаления этого осциллятора из системы? На этот вопрос следует ответить положительно, рассмотрев обратный процесс. Пусть мы добавляем в систему невырожденных осцилляторов ещё один осциллятор, собственная частота которого совпадает с частотой одного из нормальных колебаний системы. В новой системе эта частота, очевидно, перестанет быть частотой нормальных колебаний, что очевидно из Рис.8. Но обратный процесс удаления данного осциллятора опять делает эту частоту частотой нормальных колебаний.

На частотах остальных осцилляторов амплитуда возбуждаемого осциллятора принимает значения:

$$u_i(\omega_j) = \frac{f_i}{m_i}\frac{1}{\omega_i^2 - \omega_j^2} \qquad j \neq i \qquad (33)$$

знак которых зависит от разности частот осцилляторов. Амплитуда колебаний осциллятора с $j \neq i$ показана на рис.7с, из которого следует, что амплитуды таких осцилляторов никогда не обращаются в нуль и принимают на своих собственных частотах следующие значения:

$$u_j(\omega_j) = -\frac{f_i}{m_j c_j}\frac{c_i}{\omega_i^2 - \omega_j^2} \qquad (34)$$

Поведение амплитуд осцилляторов с частотами $\omega_j$ $j \neq i$ вблизи частот нормальных колебаний аналогично по виду поведению амплитуды тележки, но фаза их движения по отношению к действующей силе зависит дополнительно от знака величины $(\omega_j^2 - \omega^2)$.

## 6.4 Одновременное возбуждение тележки и осциллятора

Обратимся, наконец, к случаю, когда на одной и той же частоте возбуждаются как тележка, так и осциллятор с собственной частотой $\omega_i$. Именно этот случай представляет особый интерес, поскольку, как это будет видно из дальнейшего, именно эта классическая задача имеет некоторое сходство с квантово-механической задачей Фано. Уравнения принимают при этом вид:

$$(\omega_i^2 - \omega^2)u_i - c_i U = \frac{f_i}{m_i} \tag{35}$$

$$(\omega_j^2 - \omega^2)u_j - c_j U = 0 \qquad j \neq i \tag{36}$$

$$(\Omega^2 - \omega^2)U - \sum_{i=-N}^{N} \frac{m_i}{M} c_i u_i = \frac{F}{M} \tag{37}$$

а их решения выглядят следующим образом:

$$U = \frac{1}{M} \frac{F + f_i \frac{c_i}{\omega_i^2 - \omega^2}}{(\Omega^2 - \omega^2) - \sum_{k=-N}^{N} \frac{m_k}{M} \frac{c_k^2}{\omega_k^2 - \omega^2}} \tag{38}$$

$$u_i = \frac{1}{\omega_i^2 - \omega^2} \left\{ \frac{c_i}{M} \frac{F + f_i \frac{c_i}{\omega_i^2 - \omega^2}}{(\Omega^2 - \omega^2) - \sum_{k=-N}^{N} \frac{m_k}{M} \frac{c_k^2}{\omega_k^2 - \omega^2}} + \frac{f_i}{m_i} \right\} \qquad \omega \neq \omega_i \tag{39}$$

$$u_j = \frac{1}{\omega_j^2 - \omega^2} \frac{c_j}{M} \frac{F + f_i \frac{c_i}{\omega_i^2 - \omega^2}}{(\Omega^2 - \omega^2) - \sum_{k=-N}^{N} \frac{m_k}{M} \frac{c_k^2}{\omega_k^2 - \omega^2}} \tag{40}$$

Новым, по сравнению со случаем возбуждения только одного осциллятора, является появление в графиках новых нулей и изменение положения старых как показано на рис.9. Новый нуль амплитуды тележки даётся выражением:

$$F + f_i \frac{c_i}{\omega_i^2 - \omega^2} = 0 \tag{41}$$

Из уравнения (41) находим:

$$\omega_U = \sqrt{\omega_i^2 + \frac{f_i}{F} c_i} \tag{42}$$

Этот нуль существует при разумном отношении сил всегда, поскольку в нашей модели $c_i \ll \omega_i^2$, однако в зависимости от знака этого отношения может быть как $\omega_U < \omega_i$ так и $\omega_U > \omega_i$. Значение $U(\omega_i) = -\frac{f_i}{m_i c_i}$ остаётся тем же, что и в случае $F = 0$ и не зависит от положения нового нуля. Такой же нуль появляется и в графике $u_j$, как это следует из выражения (40). В графике $u_i$ новых нулей не появиться, однако «старые» испытают некоторый сдвиг. Это нетрудно увидеть из уравнения для нулей $u_i$, которое приобретает теперь вид:

$$(\Omega^2 - \omega^2) - \sum_{\substack{k=-N \\ k \neq i}}^{N} \frac{m_k}{M} \frac{c_k^2}{\omega_k^2 - \omega^2} = -\frac{m_i c_i}{M} \frac{F}{f_i} \tag{43}$$

Данное уравнение совпадает с вековым уравнением системы, из которой удалён осциллятор с частотой $\omega_i$ и изменена собственная частота колебаний тележки:

$$\widetilde{\Omega}^2 = \Omega^2 + \frac{m_i c_i}{M} \frac{F}{f_i} \tag{44}$$

Взглянув на графическое решение векового уравнения на Рис.3b легко увидеть, что сдвиг собственной частоты тележки приведёт к сдвигу всех корней этого уравнения в одну и ту же сторону.

Так же, как и в предыдущем случае, нетрудно получить из исходных уравнений значение амплитуды возбуждаемого внешней силой осциллятора на его собственной частоте.

$$u_i(\omega_i) = -\frac{M f_i}{(m_i c_i)^2} \left[ (\Omega^2 - \omega_i^2) - \sum_{k \neq i} \frac{m_k}{M} \frac{c_k^2}{\omega_k^2 - \omega_i^2} \right] - \frac{F}{m_i c_i} \tag{45}$$

### 7. Предельный переход к непрерывному спектру осцилляторов

Имея достаточно подробное представление о поведении нашей консервативной системы при возбуждении её внешними периодическими силами, перейдём теперь к поиску предельного перехода к непрерывному спектру осцилляторов. Выберем частоты осцилляторов кратными некоторой основной частоте $\nu = \frac{\Omega}{2N+1}$ так, чтобы $\omega_i = (2N + 1 +$ *i*) v. Для того, чтобы не нарушалось наше исходное предположение (8) о слабости связи между тележкой и всё возрастающим количеством осцилляторов, необходимо потребовать, чтобы: $\sum_{i=-N}^{N} c_i \ll \Omega^2$. Примем, для простоты, что все коэффициенты связи равны между собой и уменьшаются с ростом числа осцилляторов в системе: $c_i = c = \frac{V}{2N+1}$, $V \ll \Omega^2$. Поскольку величина $V$ ограничена, переход к бесконечному количеству осцилляторов влечёт за собой стремление к нулю коэффициентов связи.

Решение задачи Фано даёт выражения для собственных векторов неограниченной системы с непрерывным спектром. Наша система связанных осцилляторов всегда остаётся локализованной в пространстве, однако можно попытаться, увеличивая число осцилляторов, сколь угодно приблизить спектр их частот к непрерывному. Из рис. 3 видно, что частоты нормальных колебаний нашей невырожденной системы будут располагаться по одной между частотами соседних осцилляторов, так что при увеличении плотности частот осцилляторов (количества осцилляторов на единичный частотный интервал) будет возрастать и плотность частот нормальных колебаний.

Аналогом уравнения на собственные значения энергии конечномерной задачи Фано [4] для нашей системы является вековое уравнение:

$$(\Omega^2 - \omega_\lambda^{\,2}) - \sum_{k=-N}^{N} \frac{c^2}{\omega_k^2 - \omega_\lambda^{\,2}} = 0 \qquad (46)$$

в котором мы, для простоты, положили все массы пружинных осцилляторов одинаковыми и равными массе тележки и явным образом указали частоту нормального колебания $\omega_\lambda$. В силу кратности частот для нашей системы:

$$\sum_{k=-N}^{N} \frac{c^2}{\omega_k^2 - \omega_\lambda^{\,2}} = \frac{\pi c^2}{\nu^2} \sum_{k=-N}^{N} \frac{1}{\pi(2N+1+k)^2 - \pi\left(\frac{\omega_\lambda}{\nu}\right)^2} = -\frac{\pi c^2}{\nu^2} \cot \pi \left(\frac{\omega_\lambda}{\nu}\right)^2 \qquad (47)$$

и вековое уравнение принимает вид:

$$(\Omega^2 - \omega_\lambda^{\,2}) + \frac{\pi c^2}{\nu^2} \cot \pi \left(\frac{\omega_\lambda}{\nu}\right)^2 = 0 \qquad (48)$$

Это уравнение полностью аналогично уравнению на собственные значения энергии полученному в первой работе Фано при рассмотрении взаимодействия дискретного атомного состояния с квазиконтинуумом состояний [1].

$$E = \frac{q^2 \pi}{\tau} \cot \frac{E\pi}{\tau} \qquad (49)$$

при замене: $\omega_\lambda{}^2 \to E$, $c \to q$, $\nu^2 \to \tau$ и выборе $\Omega^2$ в качестве начала отсчёта энергии, как это и сделано в работе Фано. (Здесь необходимо заметить, что символом $q$ в работе [1] обозначена величина пропорциональная матричному элементу перехода между дискретным состоянием и состояниями квазиконтинуума вполне аналогичная коэффициенту связи $c$, а не параметр резонансного контура Фано.) Величина $\nu$ -частотный шаг, а $\tau$ – энергетический шаг квазиконтинуума.

Собственным состояниям задачи Фано с заданной энергией $E$ очевидно соответствуют нормальные колебания нашей системы с заданной частотой $\omega_\lambda$. Тогда амплитуде волновой функции будет соответствовать амплитуда нормального колебания и квадрату модуля волновой функции, то есть вероятности, квадрат амплитуды нормального колебания, то есть мощность спектра. Но на квадрат модуля волновой функции наложено ограничение, связанное с нормировкой, тогда как квадрат амплитуды нормального колебания неограниченно возрастает при возбуждении системы на её резонансной частоте. В результате, частотная зависимость амплитуд колебаний осцилляторов будет содержать разрывы на частотах нормальных колебаний, несмотря на увеличение спектральной плотности числа осцилляторов. При возбуждении такой системы на частотах нормальных колебаний будет наблюдаться неограниченный рост амплитуды колебаний, тогда как при сколь угодно малом отступлении от этой частоты амплитуда колебаний будет конечной. Таким образом, амплитуда колебаний осцилляторов не стремится к конечному пределу при неограниченном повышении спектральной плотности нормальных колебаний. Введение малого затухания для осцилляторов кардинально меняет ситуацию и переход к пределу непрерывного спектра становится возможным. Затухание в нашей классической модели играет ту же роль, которую играет нормировка волновой функции в квантовой.

В случае континуума частот возбуждение всегда происходит на собственной частоте состояния, входящего в континуум. Если возбуждать систему только на частоте, совпадающей с одной из собственных частот входящих в неё осцилляторов (уравнения (35) - (37) с $\omega = \omega_i$) и интересоваться, при этом, амплитудой колебания возбуждаемого осциллятора (45), то

$$u_i(\omega_i) = -\frac{M f_i}{(m_i c_i)^2}[\Omega^2 - \omega_i^2 + R(\omega_i)] - \frac{F}{m_i c_i} = u_{0c}(q + \epsilon) \qquad (50)$$

Здесь

$$R(\omega_i) = -\sum_{k \neq i} \frac{m_k}{M} \frac{c_k^2}{\omega_k^2 - \omega_i^2} \qquad (51)$$

и коэффициенты связи предполагаются слабо зависящими от частоты $\omega_i$, так что $R(\omega_i) \neq 0$. В формуле введены величины аналогичные величинам, используемым в работе Фано [2]:

$$\epsilon = \frac{1}{c_i}[\omega_i^2 - \Omega^2 - R(\omega_i)] \qquad (52)$$

$$q = -\frac{F}{f_i}\frac{m_i}{M} \qquad (53)$$

и амплитуда возбуждения состояний континуума: $u_{0c} = \frac{M}{m_i^2} \frac{f_i}{c_i}$. В итоге получаем:

$$\left(\frac{u_i(\epsilon)}{u_{0c}}\right)^2 = (q + \epsilon)^2 \qquad (54)$$

результат эквивалентный числителю формулы Фано. Мы видим, что при условии

$$q + \epsilon = 0 \qquad (55)$$

возбудить колебания осциллятора не удаётся. Этот результат совпадает с условием минимума формулы Фано.

Покажем, что при введении затухания осцилляторов квазиконтинуума удаётся получить также и резонансный знаменатель формулы Фано. Для этого введём в выражение (39) феноменологическое затухание $\gamma$ одинаковое для всех пружинных осцилляторов:

$$u_i = \frac{1}{\omega_i^2 - \omega^2 - i\gamma\omega} \left\{ \frac{c_i}{M} \frac{F + f_i \frac{c_i}{\omega_i^2 - \omega^2 - i\gamma\omega}}{(\Omega^2 - \omega^2) - \sum_{k=-N}^{N} \frac{m_k}{M} \frac{c_k^2}{\omega_k^2 - \omega^2 - i\gamma\omega}} + \frac{f_i}{m_i} \right\} \tag{56}$$

Введём безразмерную энергию $\epsilon$, как это было сделано выше. При вычислении малой поправки $R(\omega_i)$ к резонансной частоте тележки пренебрежём затуханием и будем интересоваться случаем $\omega = \omega_i$:

$$u_i(\omega_i) = \frac{1}{i\gamma\omega_i}\left\{ \frac{1}{M} \frac{F - f_i \frac{c_i}{i\gamma\omega_i}}{\epsilon - \frac{m_i}{M}\frac{c_i}{i\gamma\omega_i}} - \frac{f_i}{m_i} \right\} = u_{0c} \frac{q+\epsilon}{1 + i\frac{\gamma\omega_i M}{m_i c_i}\epsilon} \tag{57}$$

И для квадрата модуля отношения амплитуд имеем:

$$\left(\frac{u_i(\epsilon)}{u_{0c}}\right)^2 = \frac{(q+\epsilon)^2}{1 + \left(\frac{\gamma\omega_i M}{m_i c_i}\right)^2 \epsilon^2} \tag{58}$$

Таким образом, мы видим, что при введении малого затухания для осцилляторов получается приближённый результат эквивалентный по сути квантово-механической формуле Фано. Физическим смыслом затухания является учёт энергии, уходящей из системы осцилляторов на бесконечность. Если мы интересуемся только релаксацией дискретного метастабильного состояния, то можно вместо рассмотрения непрерывного спектра состояний ввести феноменологическое затухание. В расширенной постановке задачи, включающей состояния непрерывного спектра, феноменологическое затухание уже не требуется, поскольку часть системы, обеспечивающая релаксацию дискретного состояния, включена в рассмотрение явно. Как в квантовой, так и в классической задаче энергия должна уходить на бесконечность с ростом размеров системы. Именно этот уход и моделирует феноменологическое затухание осцилляторов квазиконтинуума. Это затухание должно масштабироваться с ростом числа осцилляторов подобно масштабированию коэффициентов связи. Для затухающих осцилляторов квазиконтинуума амплитуда нормальных колебаний становится конечной и при неограниченном росте спектральной плотности осцилляторов амплитуды колебаний будут стремиться к одному и тому же пределу при сближении частот. Если при масштабировании коэффициентов связи положить

$\frac{\gamma \omega_i M}{m_i c_i} = 1$, то есть $\gamma = \frac{m_i c_i}{\omega_i M}$ (59)

то затухание будет уменьшаться пропорционально уменьшению коэффициентов связи и формула (58) совпадёт с формулой Фано.

В нашей модели отсутствует часть системы, обладающая по-настоящему непрерывным спектром, то есть совершающая инфинитное движение. Для того, чтобы нормальные колебания нашей системы перешли в нормальные колебания задачи Фано недостаточно сделать сколь угодно плотной частоту нормальных колебаний, необходимо также обеспечить возможность инфинитного движения, то есть возможность ухода энергии, поступающей в систему, на бесконечность. Классической моделью, адекватной задаче Фано, мог бы, по-видимому, стать локальный осциллятор, взаимодействующий с волноводом конечной длины, которая при переходе к случаю непрерывного спектра неограниченно возрастает, что схематично показано на Рис.10. Это должно обеспечить переход нормальных колебаний, описываемых стоячими волнами, в нормальные колебания бегущих волн.

## 8. Заключение

Мы детально рассмотрели систему взаимодействующих классических осцилляторов, и показали, что при некоторых условиях она во многом, но не во всём аналогична квантово-механической системе дискретного энергетического уровня, взаимодействующего с энергетическим квазиконтинуумом состояний. Был проанализирован предельный переход к случаю непрерывного спектра и возможная связь исследуемой задачи с генерацией когерентных фононов сверхкороткими лазерными импульсами. Мы показали, что с принципиальной точки зрения введение феноменологического затухания не делает рассмотренную выше задачу полностью эквивалентной задаче Фано, хотя в практическом отношении аналогия, установленная в работе [16], может оказаться полезной. Классическая аналогия задачи Фано может иметь место, но не в механике частиц, а в механике сплошных сред, то есть волн, - это локальный осциллятор, слабо связанный с внешними бегущими волнами. За более детальным ознакомлением с отдельными деталями Фано резонанса мы отсылаем читателя к цитированным выше статьям, обзорам и монографиям.

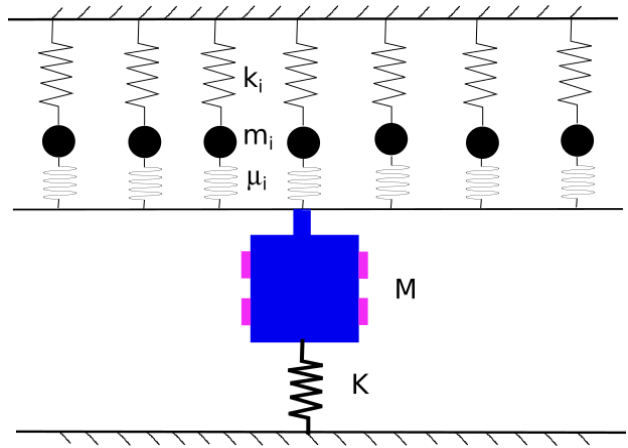

Рис. 1. Схематическое изображение связанных гармонических осцилляторов (вид сверху). На горизонтальной плоскости расположена тележка массы $M$, которая может двигаться по рельсам под действием пружины жесткостью $K$, связанная слабыми пружинами с набором пружинных осцилляторов.

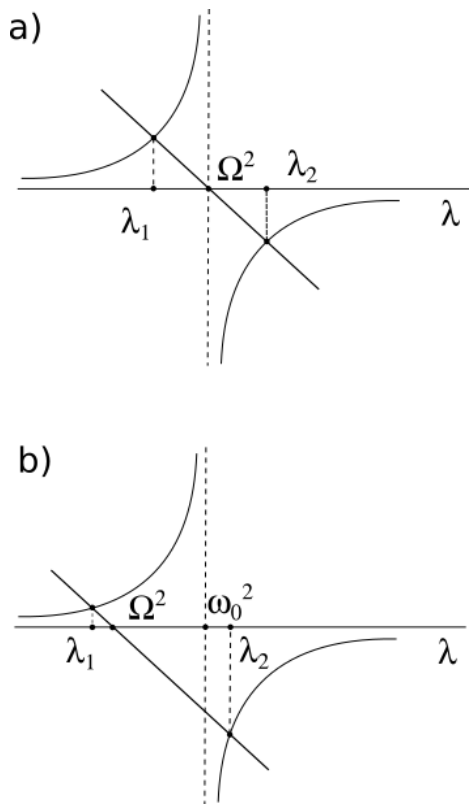

Рис. 2. Графическое решение векового уравнения для случая двух связанных осцилляторов (пружинный маятник + тележка). (a) - частоты пружинного маятника и тележки совпадают, (b) -эти частоты различны.

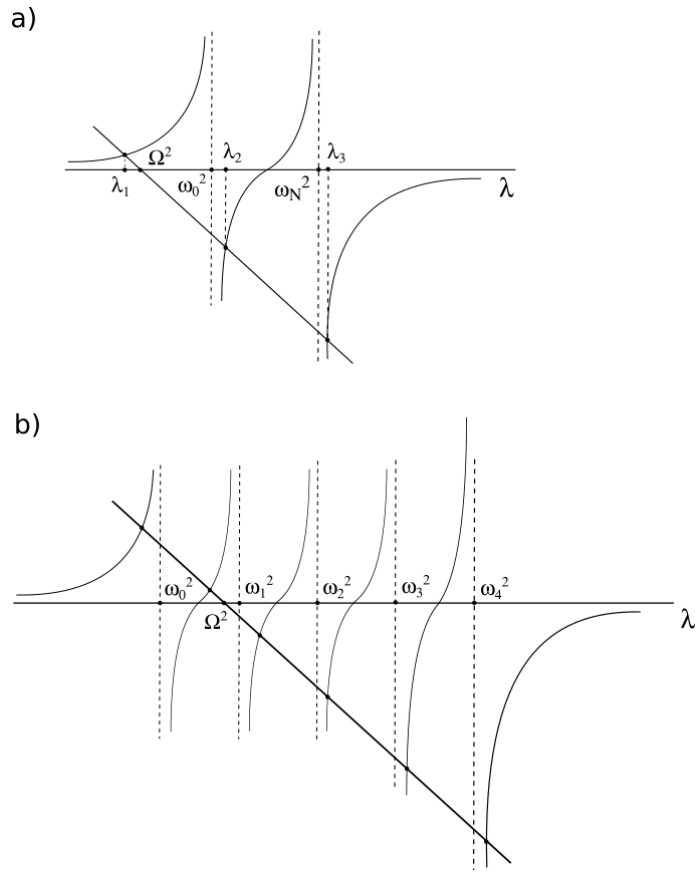

Рис.3. Графическое решение векового уравнения при снятии вырождения: (a) для случая, когда собственная частота $\omega_0$ одного из осцилляторов отличается от частоты собственных колебаний тележки $\Omega$ и от собственной частоты остальных $2N$ осцилляторов $\omega_N$, (b) – все частоты собственных колебаний маятников различны.

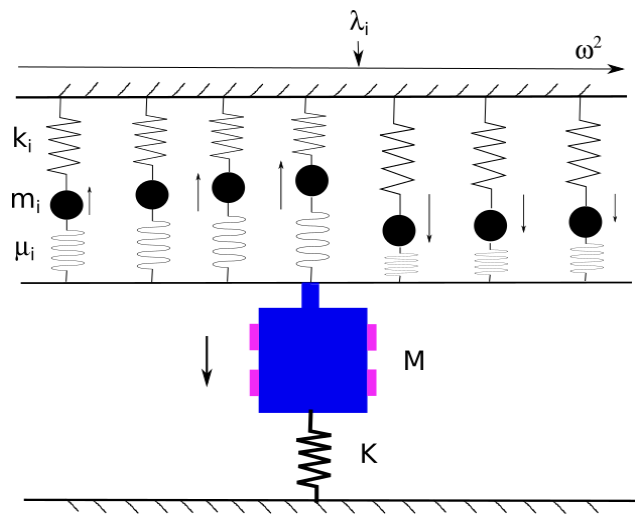

Рис. 4. Наглядное представление движения пружинных маятников при возбуждении нормального колебания $\lambda_j$: маятники с собственными частотами $\omega_i^2 > \lambda_j$ движутся синфазно, а маятники с $\omega_i^2 < \lambda_j$ - противофазно с тележкой.

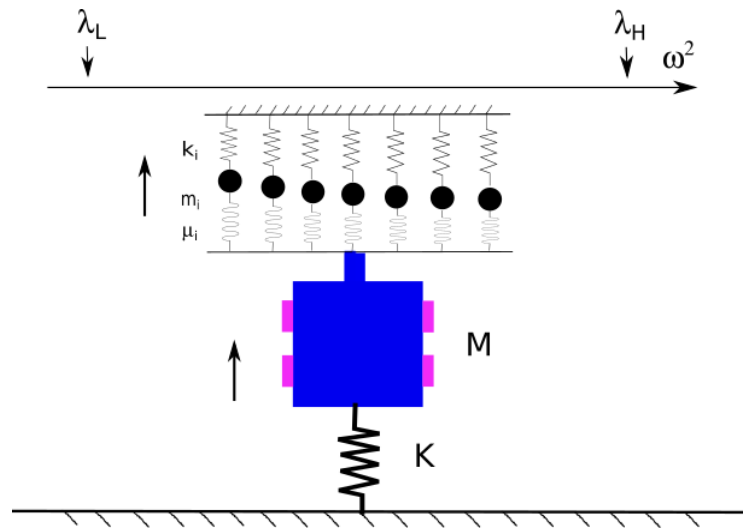

Рис.5. Движение пружинных маятников и тележки при возбуждении синфазного граничного колебания $\lambda_L$.

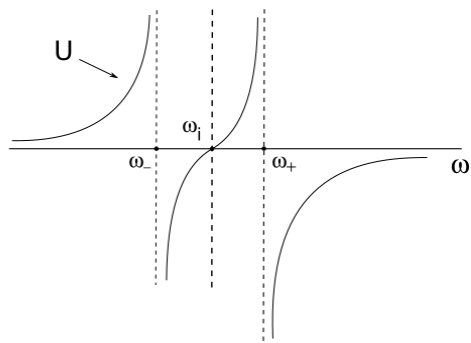

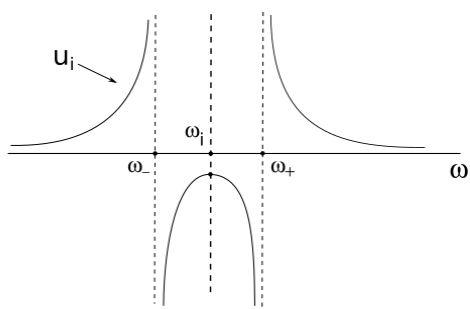

Рис.6. Зависимость амплитуды колебаний (а) – тележки и (в) – одного из пружинных маятников от частоты гармонической вынуждающей силы, изменяющейся в окрестности его резонансной частоты $\omega_i$.

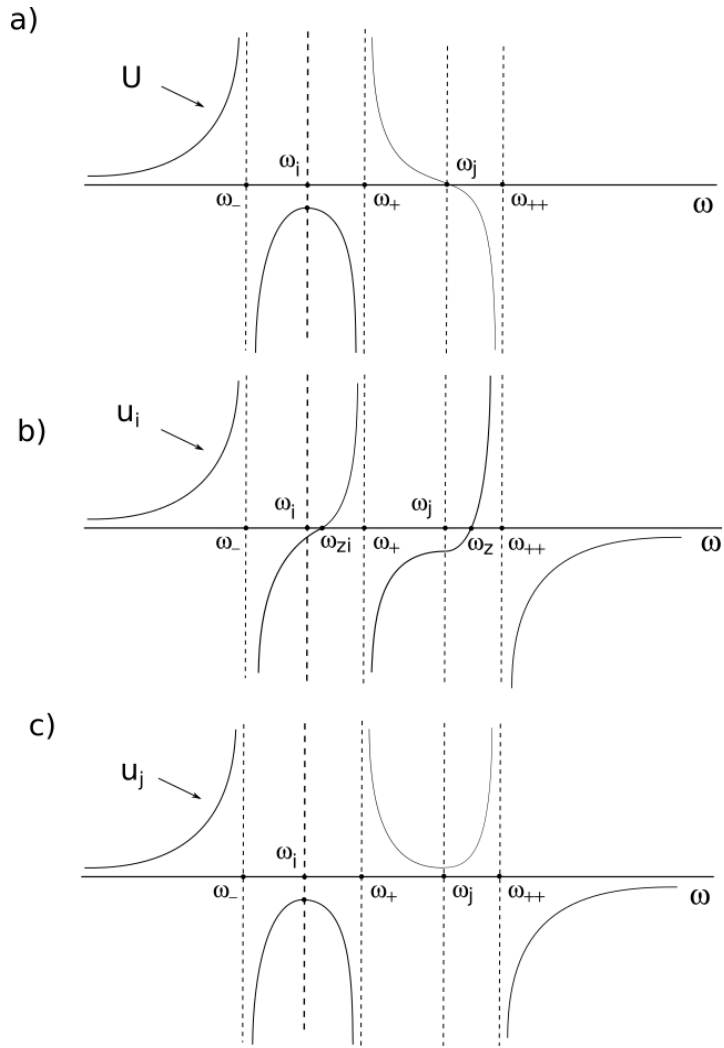

Рис.7. Частотная зависимость амплитуд колебаний пружинных осцилляторов и тележки при действии гармонической силы только на осциллятор с собственной частотой $\omega_i$. Частоты $\omega_-$, $\omega_+$ и $\omega_{++}$ соответствуют нормальным колебаниям, $\omega_z$ и $\omega_{zi}$ – частоты нормальных колебаний «укороченной» системы осцилляторов (см. текст).

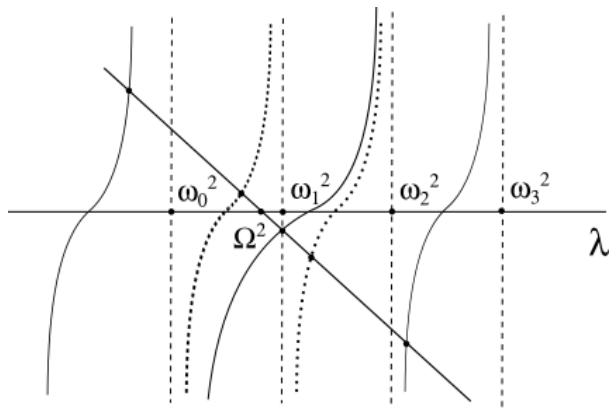

Рис.8 Графическое решение векового уравнения для невырожденной системы, состоящей из тележки и осцилляторов с частотами $\omega_0$, $\omega_2$ и $\omega_3$ (сплошные линии) изменяется после добавления в систему ещё одного осциллятора с частотой $\omega_1$, совпадающей с частотой одного из нормальных колебаний (пунктирные линии). При этом $\omega_1$ перестаёт быть частотой нормального колебания системы. Обратный процесс удаления из системы осциллятора с частотой $\omega_1$ делает его частоту частотой нормального колебания. Небольшое изменение частот остальных нормальных колебаний, отстоящих от частоты $\omega_1$ на рисунке не показано.

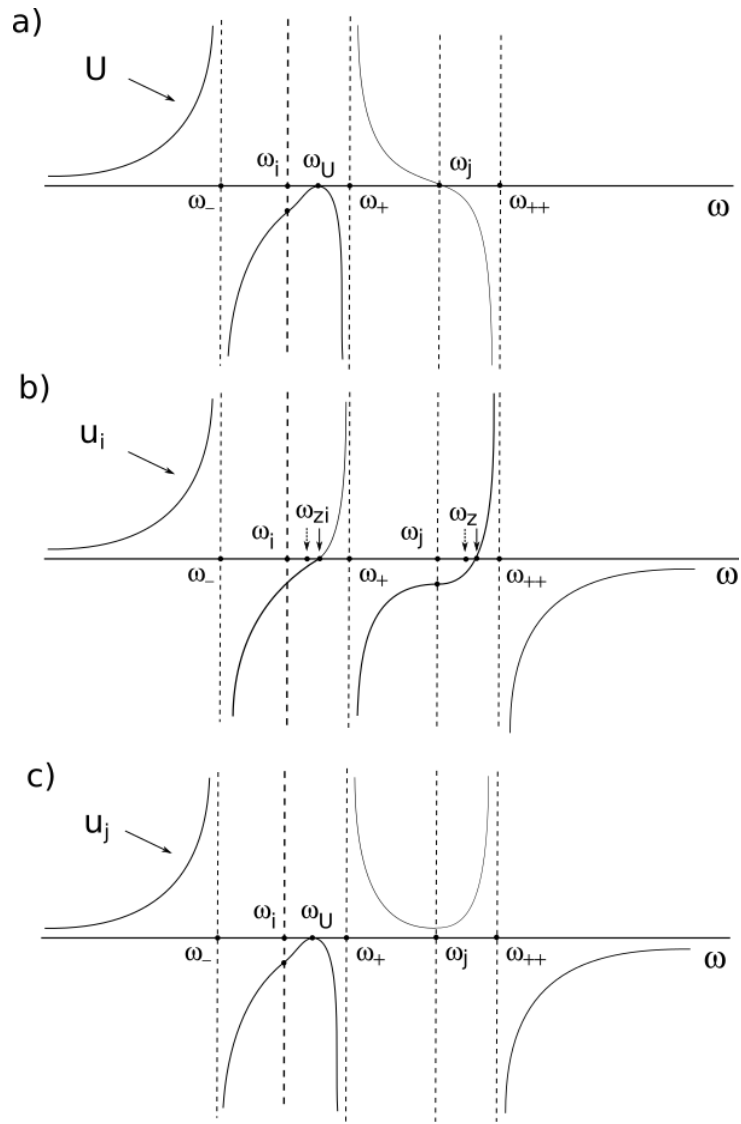

Рис. 9. Одновременное возбуждение тележки и осциллятора с собственной частотой $\omega_i$ гармоническими силами с одной и той же фазой и частотой ω. По сравнению с рис.7 в окрестности $\omega_i$ появляется новый нуль амплитуды тележки и остальных осцилляторов и изменяется положение нулей осциллятора $\omega_i$.

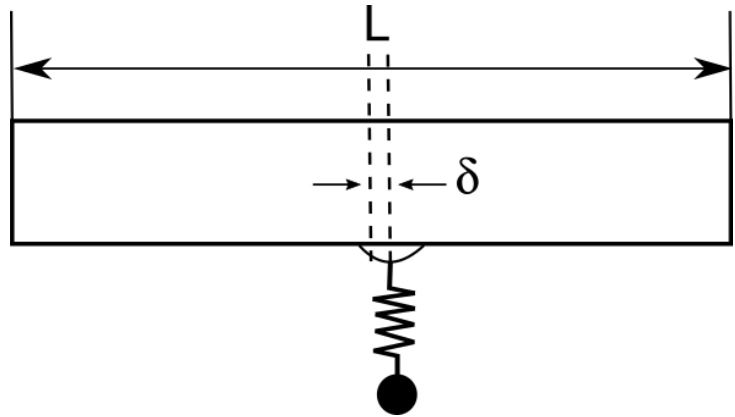

Рис.10. Классическая колебательная система, состоящая из акустического резонатора длины L, в котором могут возбуждаться звуковые колебания, и локального пружинного осциллятора, связанного с резонатором посредством гибкой мембраны. При возрастании L система сможет обеспечивать уход энергии на бесконечность оставаясь консервативной.